\documentclass{article}

\usepackage{arxiv}

\usepackage[utf8]{inputenc} 
\usepackage[T1]{fontenc}    
\usepackage{hyperref}       
\usepackage{url}            
\usepackage{booktabs}       
\usepackage{amsfonts}       
\usepackage{nicefrac}       
\usepackage{microtype}      
\usepackage{lipsum}
\usepackage{graphicx}

\usepackage[utf8]{inputenc} 
\usepackage[T1]{fontenc} 
\usepackage{multirow} 

\usepackage{multicol} 

\usepackage{rotating} 

\usepackage{float} 

\usepackage{comment} 

\usepackage{wrapfig} 

\usepackage[inline]{enumitem} 

\usepackage[normalem]{ulem}
\useunder{\uline}{\ul}{}

\usepackage{caption}
\usepackage{subcaption} 
\usepackage[table,xcdraw]{xcolor}

\graphicspath{ {./Figures/} }

\title{Towards Smart Networking with SDN Enabled IPv6 Network}

\author{
 Babu R. Dawadi \\
  Department of Electronics and Computer Engineering, \\
  Institute of Engineering, Pulchowk Campus, \\
  Tribhuvan University, Nepal \\
  \texttt{baburd@ioe.edu.np} \\
   \And
 Shashidhar R. Joshi \\
  Department of Electronics and Computer Engineering, \\
  Institute of Engineering, Pulchowk Campus, \\
  Tribhuvan University, Nepal \\
  \texttt{srjoshi@ioe.edu.np} \\
  \And
 Danda B. Rawat \\
  Cyber Security and Wireless Networking Innovations Lab, \\
  EECS department, Howard University, \\
  Washington, DC 20059, USA \\
  \texttt{db.rawat@ieee.org} \\
  \And
  Pietro Manzoni \\
  Department of Computer Engineering, \\
  Universitat Politècnica De València, \\
  46022 Valencia,~Spain \\
  \texttt{pmanzoni@disca.upv.es} \\
}

\begin{document}
\maketitle
\begin{abstract}
This paper presents the features and benefits of legacy IPv4 network migration towards major two latest networking paradigms viz. Internet protocol version 6 (IPv6) and the software-defined networking (SDN). These latest networking paradigms are the enabler of future generation networking so that the standards and requirements of fifth generation (5G) wireless networking can be achieved. Features and migration approaches of IPv6 and SDN will be separately discussed, then a joint migration approach of SDN and IPv6 network termed as SoDIP6 network migration will be presented, and the integration of SoDIP6 network as a backbone of 5G network will be introduced. 
\end{abstract}


\section{Introduction and Background}
The rapid increase of internet users Worldwide with the increased Internet of Things (IoT) smart devices, and the trend of World moving to converged network environment into the mode of computer networks (IP based network) encourage researchers, developers, and the networking enterprises Worldwide to enhance the intelligence in networking technologies by moving to next generation networking paradigm viz. IPv6 addressing mechanism, Software-Defined Networking (SDN), network function virtualization (NFV), wireless sensor networks (WSN), and towards the complete digital packet based next generation communications networks. 

The tremendous growth of ICT based services and businesses in the World made the early exhaustion of IPv4 addressing and compelled us to implement standardized new addressing scheme called Internet protocol version 6 (IPv6) that has sufficient address space with 128 bit length. In the current and future generation networking, every connected device have more than one unique IPv6 addresses no matter how much the internet users exist in the long future because IPv6 address space is more than astronomical. 

The shortage of IPv4 addresses enforces the networking organizations World-wide in the phase of transition to IPv6 addressing mechanism by following best migration approach with optimum cost of migration. Additionally, the existing legacy network has identified with several problems including security, quality of service, routing, device configurability, monitoring, control, and the major thing is vertically integrated having control and data plane bundled on each switches and routers.  Meanwhile, the SDN, a new concept in network operation and management, has been introduced with the considerable changes in networking paradigms by introducing open standards and enabling the programmable network with the segregation of data and control plane for efficient network management so that its implementation with IPv6 addressing enables towards achieving requirements standardized by fifth generation (5G) wireless communications.

IPv6 on the one hand improves the efficiency of internet protocol as a whole including routing, while on the other hand SDN improves the controllability of networking equipment through programming and virtualization with an approach to using open protocols, such as Open-Flow, to apply globally aware software control at the network's edges to access network devices that would normally use closed and vendor specific firmware.

\begin{figure}[ht!]
    \centering
    \includegraphics[width=0.8\textwidth]{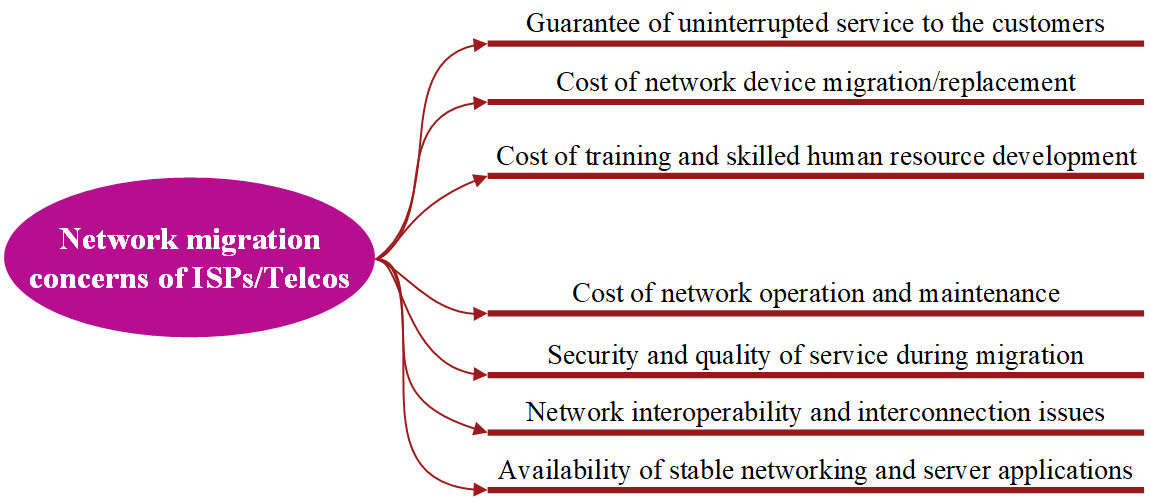}
    \caption{Network migration concerns for ISPs and Telcos}
    \label{fig:NetMigrationConcern}
\end{figure}

The invention of those new concepts and development creates a bigger challenges in networking for service providers to migrate their existing legacy networks into the software defined based IPv6 enabled network. The major concerns of network migration for the internet and telecom service providers are depicted in Figure \ref{fig:NetMigrationConcern}.

The convergence in telecommunications focuses towards the migration to latest generation digital packet based communications leading to the requirements of software controlled IPv6 network against the already exhausted IPv4 addresses. But immediate migration of the existing network into SDN and IPv6 based networking is not possible. The migration is a gradual process that necessitates the development of a proper strategy that takes into account technology requirements, customer demand, capital expenditure (CapEX), operational expenditure (OpEX), and traffic engineering views in order to ensure a smooth transition. The transition phase may last longer, during when service providers must migrate their network using the best migration strategy and at the lowest possible cost.

Basically service providers of developing countries are facing challenges of network migration due to the lack of sufficient funds in addition with lack of skilled human resources to manage and operate the new technologies. Hence, a requirement on techno-economic analysis for proper migration of existing network into SDN based IPv6 network is the need to develop smart future networks.

The term ``migration'' refers to the process of upgrading an existing network to make it compatible with new technology. The major components of the service network that can operate with new technologies are routers and switches. With the investment in CapEX and OpEX for service providers, either a software/firmware upgrade or device replacement is required. This paper provides the overview of IPv6 addressing and SDN with the need of joint network migration for smooth and cost effective transition to latest smart networking.

Cost estimation of network device migration is a complex task because of the dynamism in equipment's specification and vendor specific configuration and management. Migration cost modeling shall be performed with statistical, economic assessment, and intelligent approaches. 

\begin{figure}[ht!]
    \centering
    \includegraphics[width=0.5\textwidth]{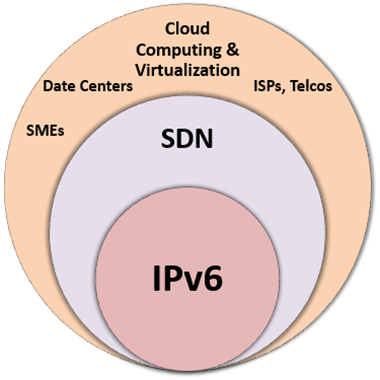}
    \caption{Latest networking paradigms}
    \label{fig:FutureNetworkParadigm}
\end{figure}

Three sectors in the networking and virtualization domain as depicted in Figure \ref{fig:FutureNetworkParadigm} are the major inventions that the world's network and cloud service providers have to move on with to achieve smart of everything in the ubiquitous networking. These are: (i) IPv6 addressing mechanism, (ii) Software-Defined Networking, and iii) Cloud Computing and virtualization. Among which, cloud computing is the bigger domain which includes the networking infrastructures and platforms in its service.

\begin{figure}[ht!]
    \centering
    \includegraphics[width=0.6\textwidth]{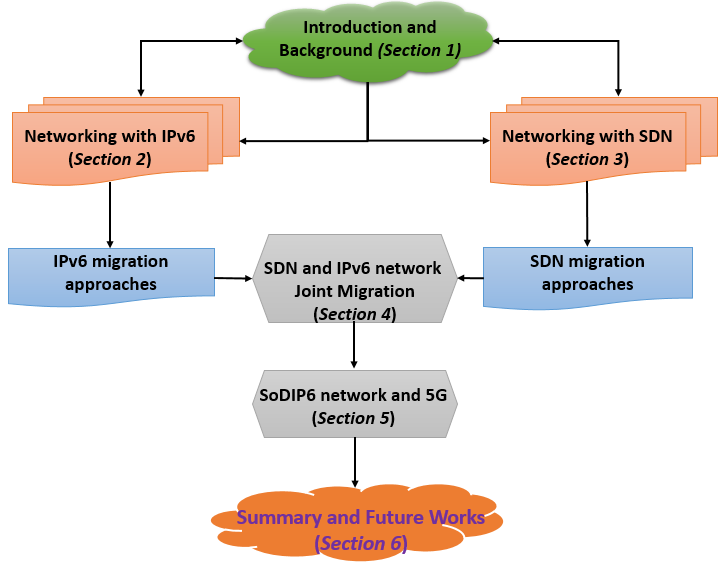}
    \caption{Structure of this paper}
    \label{fig:chapterStructure}
\end{figure}

The rest of this paper is organized as shown in Figure \ref{fig:chapterStructure}. IPv6 addressing, features and benefits with migration approaches will be discussed in section \ref{sec:IPv6Section}. Similarly, introduction of SDN with its features, implementation and migration methods will be discussed in section \ref{sec:smartNetwithSDN}. Section \ref{sec:jointSDNIPv6} introduces Joint approach of IPv6 and SDN implementation as a SoDIP6 network, while section \ref{sec:5GSoDIP6} introduces 5G concepts and its integration with SoDIP6 networks. Section \ref{sec:Summary} summarizes the paper.

\section{Networking with IPv6}
\label{sec:IPv6Section}
\subsection{Introduction and Definition}
Over the seven layers of OSI reference model, network layer is the third layer that deals with traffic routing and optimal path finding. Internet protocol (IP) is the layer 3 protocol. The introduction of Internet protocol by Vinton Cerf in early 1980s become more popular after the invention of World Wide Web (WWW) introduced by Tim Berners-Lee in early 1990s \cite{niver2016tim}. Then IP has been implemented with the introduction of 32 bit length IP address termed as IP version 4 (IPv4). The 32 bits identity number is an IP address uniquely assigned to a device, can be accessible anywhere from the World. With the introduction of IoT smart devices in which every object in this World is communicable and controlled remotely if unique IP address is assigned to it, the IPv4 address space is already exhausted that all the regional internet registries (RIR) address pools were already exhausted \cite{prehn2020wells}. In the legacy IPv4 addressing system, the networking infrastructure, hardware, and software systems with applications are developed under the compatible version of IPv4. But the network expansion has become the bottleneck due to IPv4 address depletion. Taking it as a major issue, 128 bit length Internet protocol version 6 (IPv6) has been introduced in 1998 to replace IPv4 and its associated issues in the networking World. With the introduction of IPv6 addressing, current World does not need any other new version for the next trillions of years. The higher than astronomical value of address space with IPv6 is expected to be sufficient to establish communications with everything in the other planets of this universe.

\subsection{Need and benefits of IPv6}
In this section, we comparatively discuss the IPv4 and IPv6 addressing paradigm with respect to their benefits and drawbacks, security and quality as well as routing issues with their future prospects. 

The existing IPv4 addressing infrastructure can’t solve the requirements of exponentially growing internet users in the smart network World-wide with the emergence of IoT and wireless sensor network (WSN). Table \ref{Tab:v4v6comparison} provides the complete figure of comparison between IPv4 and IPv6 under different parameters that exist as issues in IPv4 and its remedy as features in IPv6 networking. 


\begin{table}[ht!]
    \centering
    \caption{Comparison of IPv4 and IPv6}
    \begin{tabular}{c}
         \includegraphics[width=0.95\textwidth]{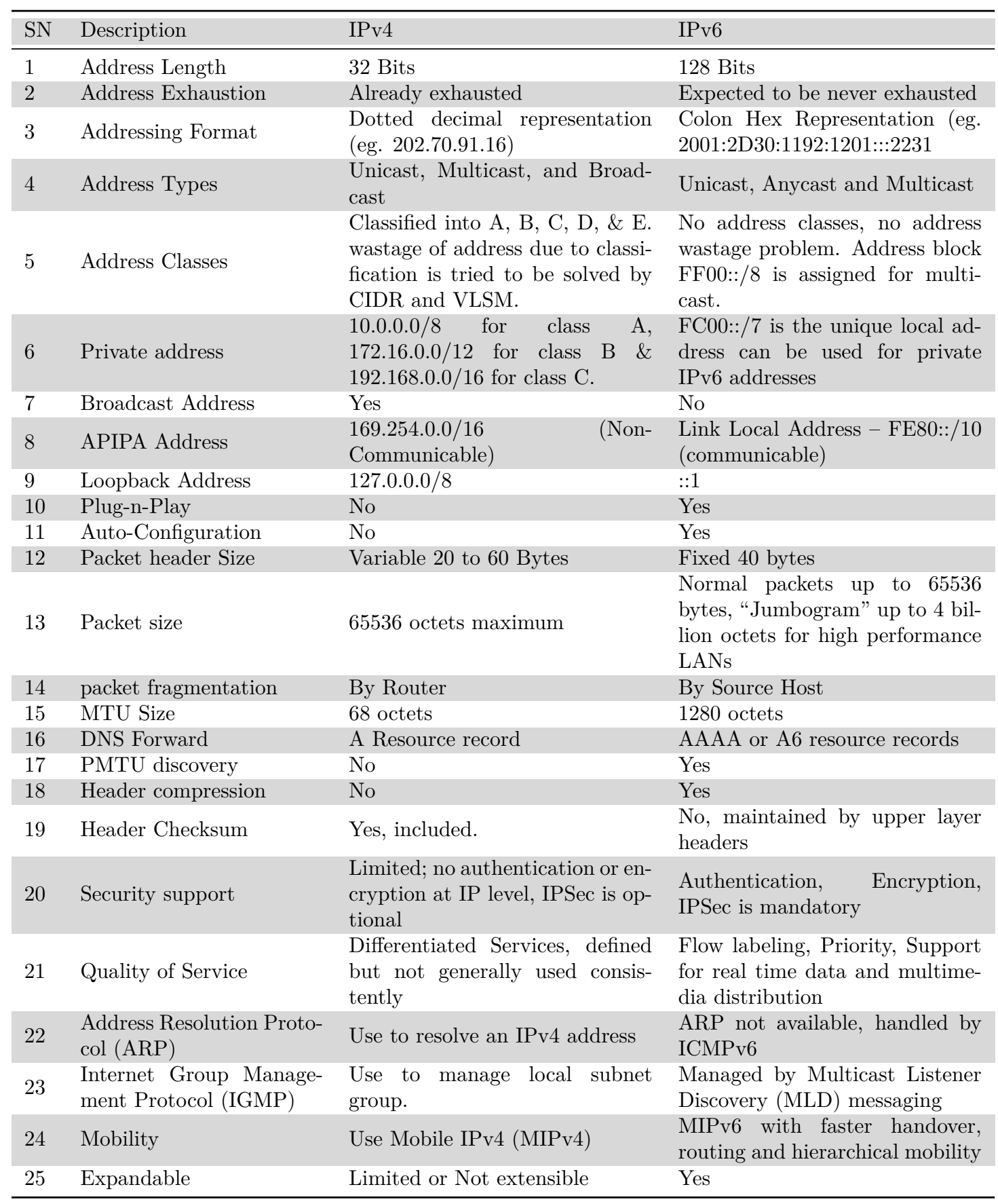}  \\
    \end{tabular}
    \label{tab:v4v6comparison}
\end{table}

\begin{figure}[ht!]
    \centering
    \includegraphics[width=\textwidth]{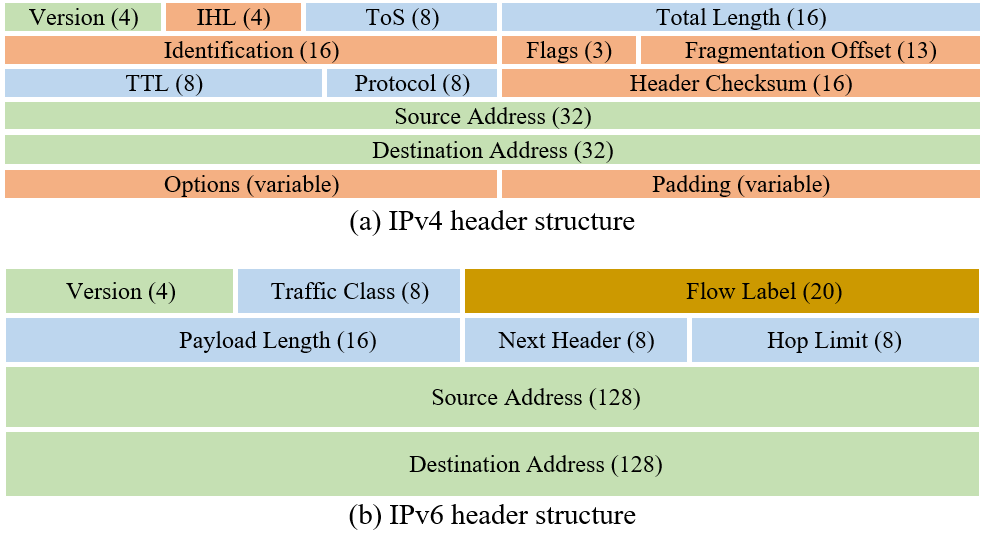}
    \caption{IPv4 and IPv6 header structure}
    \label{fig:IPv4v6HeaderStructure}
\end{figure}

IPv4 addressing and routing has several drawbacks, which are avoided in IPv6 network. Address exhaustion, network address translation (NAT) proliferation, routing table overflows, administrative hurdles due to lack of address auto-configuration, and limited expandable features including security and quality of service (QoS) issues are the major problems associated with IPv4 networks. Looking into the header structure, though IPv6 header is double the minimum header size of IPv4, it is fixed 40 bytes length in which router processing overhead is minimized by removing unnecessary header fields applied in IPv4. Figure \ref{fig:IPv4v6HeaderStructure} shows the header structure of IPv4 and IPv6. This shows that IPv6 header is more compact with less number of fields as compared with IPv4. IPv6 has introduced \emph{Flow Label} as the new header field to handle prioritized packet flow for better quality of service (QoS). Out of 40 bytes fixed header size in IPv6, the source and destination address covers 32 bytes, while rest of 8 bytes are used by router for further processing. The blue color fields in the IPv6 are modified version of IPv4 header fields as shown in Figure \ref{fig:IPv4v6HeaderStructure}. Similarly, red color fields are removed in IPv6, while green color fields are kept as it is. Flow label is newly added field in IPv6 for prioritized packet handling.

\begin{figure}[ht!]
    \centering
    \includegraphics[width=\textwidth]{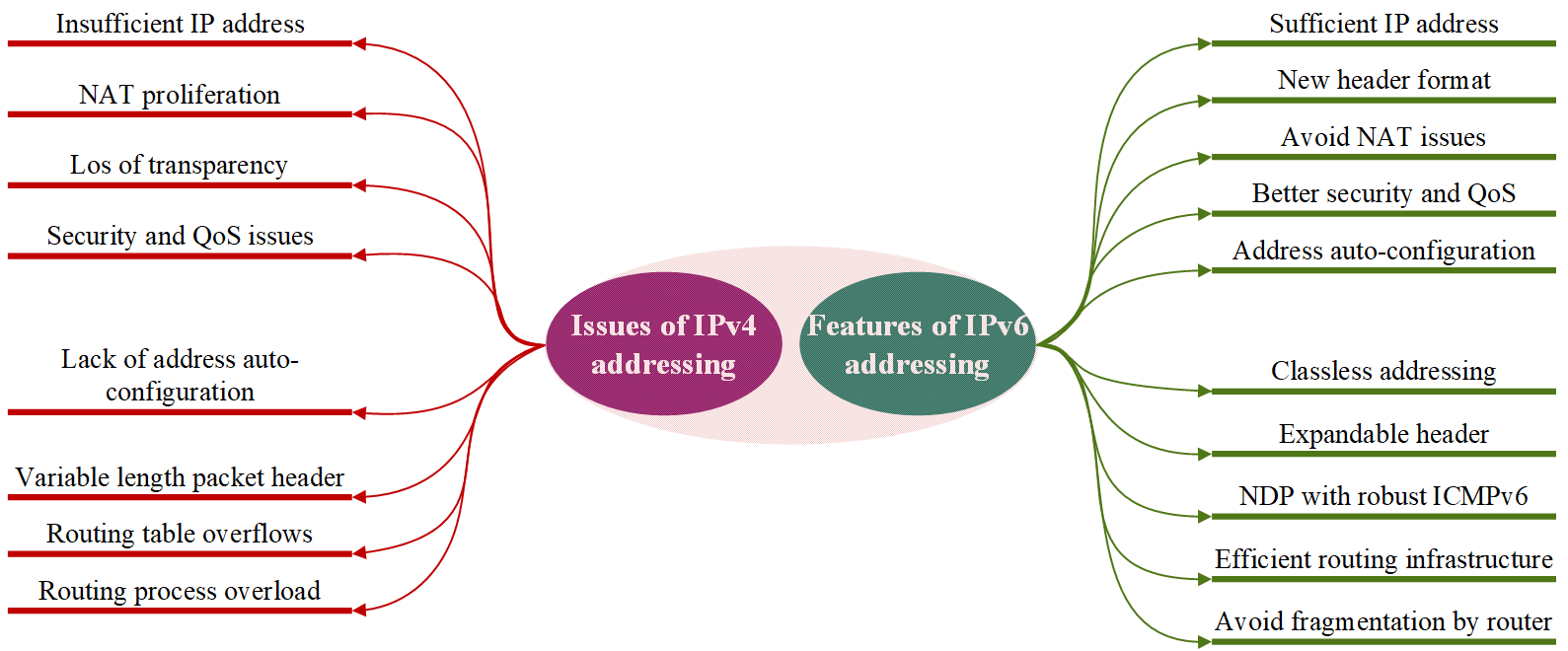}
    \caption{Issues of IPv4 and features of IPv6 addressing}
    \label{F2:ipv4ipv6issues}
\end{figure}

The functional operation of IPv6 was started in 1998 with its draft standards released by Internet engineering task force (IETF) \cite{bradner1995rfc1752}. Before entering into migration perspectives of IPv6, it is required to review on \textit{“why such migration required?”} for this, the problems of IPv4 and features of IPv6 addressing have to be analyzed so that migration to IPv6 network is justifiable. The major issues of IPv4 and features of IPv6 addressing are presented in Figure \ref{F2:ipv4ipv6issues}.

IPv4 public addresses have already been exhausted \cite{prehn2020wells}. This limits the new networks expansion for service providers using public IPv4 addresses. Besides, the ISPs who migrated their networks into IPv6 only networks, released the public IPv4 addresses, which is opened for address trading \cite{Dawadi2019,giotsas2020first} in the market. This leads to higher costs for those ISPs whose infrastructure still need to run with IPv4 addressing. The lack of public IPv4 addresses leads to obtain the address from black market, if the service providers still rely on the legacy IPv4 networking to provide the services. Alternatively, service providers are using private IPv4 addresses recursively using NAT to avoid the issues of IPv4 address depletion. But NAT has several drawbacks that it hides several computers in private zone. Private computers are not reachable via the public scope leading to transparency and administration problem. Additionally, NAT breaks the end-to-end communication due to which some applications like file transfer protocol (FTP) does not work. Most likely, NAT fails in translating embedded IP addresses and leads to application failure. As a result, IPv4 and NAT are not a viable long-term solution for service providers. On a vintage IPv4-based network, IP security (IPSec) is not mandatory in the IP layer. As a result, IPv4 security is limited. At the IP level, there is no authentication or encryption method, hence IPv4 is reliant on higher-level protocols. As a result, denial-of-service (DoS) and deception or spoofing attacks are possible. Encryption is required for packets sent at the IP level to prevent private data from being seen or manipulated. The type of service (TOS) field in the header determines QoS in IPv4. Although the Quality of Service (QoS) in IPv4 is defined, it is not consistently followed.

IPv4 lacks stateless address auto-configuration. Addressing each client personal computer (PC) can be done either manually (static addressing) or using stateful auto-configuration, for example, using dynamic host configuration protocol (DHCP) in IPv4 addressing. The static addressing in IPv4 based machines enabled only manual addressing, which, in the large network of an enterprise, is really a headache for network administrator to configure addressing over individual machines. The stateful addressing i.e. automatic address provisioning using DHCP provides the IPv4 address to a machine only to specified lease period. Major drawback of stateful configuration is that the machine might have chances to renumber its address after the lease period is expired or the machine restarts, which creates issues in proper tracking of the machine by network administrator.

IPv4 packet header length varies from 20 to 60 bytes. In the worst case, router need to process 52 bytes except the 8 bytes source and destination addresses. Dynamic header length creates burden in routing processing. Additionally, IPv4 router fragments the oversized packets by itself. Fragmentation is an extra job that a router performs for every packets when the size is greater than the maximum transmission unit (MTU) and also re-assembles by the destination router. This creates processing overhead on router due to extra task besides routing. The implementation of address aggregation \cite{Das2013} has somehow reduced the issues of IPv4 routing table overflows, however, more than 95k entries is not effective in routing information management, which still has routing table overflows in the core router in the internet.

The 128 bit length IPv6 address is designed to overcome all the issues related with IPv4 addressing. Sufficiency of IPv6 addresses led to establish internet of everything (IoE) in the universe. IPv6 avoids NAT and related issues, as every device can have many globally unique IPv6 addresses. IPv6 header is well managed, at which router only processes 8 bytes of header fields keeping 32 bytes of source and destination addresses fixed.

The fragmentation related fields available in IPv4 header are removed from the IPv6 main header so that IPv6 router never fragments the oversized packed by itself. The routing process overloading problem is avoided that the role of fragmentation is transferred to source host. Features addition in IPv6 is fairly easy. It supports extension headers after its main header that can support several optional headers to be managed in a daisy chain fashion. The size of extension headers in IPv6 is only constraints by its packet size unlike IPv4 has 40 bytes optional header field. Stateless address auto-configuration (SLAAC) in IPv6 simplifies the host configuration by enabling automatic address configuration for the link called IPv6 link-local address and derived from the prefixes advertised by a router \cite{tseng2017decision}. IPv6 enabled devices are plug-n-play as they automatically configured addresses in a link and establish communication without manual configuration. 

IP security (IPSec) support is mandatory in IPv6 that provides a standard-based solutions for security as well as promotes interoperability in different IPv6 based implementations. The dedicated flow-label field in IPv6 header provides a special handling of packets belonging to a flow, a series of packets between source and destination, enabled a prioritized delivery of packets. As compared with Internet control message protocol (ICMP) version v4, ICMPv6 is robust to monitor network health. The neighbor discovery protocol (NDP) features in IPv6 consists of several ICMPv6 messages like router/network advertisement, ICMP redirect, and multicast communications. IPv6 NDP replaces the address resolution protocol (ARP) and Internet group management protocol (IGMP) of IPv4. IPv6 addresses are categorized into unicast, anycast, and multicast. The concept of broadcast addressing of IPv4 is incorporated in the IPv6 multicast addressing.

In summary, IPv6 has stunning features and improvement as compared with the IPv4 addressing \cite{singala2018current}. The address shortage problem with associated issues of IPv4 enable service providers and enterprises World-wide to migrate their existing IPv4 network into operable IPv6 network.

\subsection{IPv6 addressing overview}
This sub-section briefly presents the addressing structure of IPv6 currently defined under IETF specification \cite{deering1998internet}. Communication in IPv6 is mainly categorized into (i) Unicast, (ii) Multicast, and (iii) Anycast. Unicast communication is for unique source to unique destination. Multicast consists of set of communication interfaces in which packets are delivered to many interfaces at a time. It is a one to many or one to all communication concept that covers the broadcast and multicast concepts of IPv4.  Anycast is defined as the set of interfaces in which packets are delivered to the nearest interface. Anycast is a kind of unicast communication but packets are delivered to all the interfaces with the concept of delivery to nearest interface first. Basically, Anycast concept is implementable to DNS query forwarding. Anycast address can be anyone unicast address that can be assigned to router only. The details of IPv6 address assignment is presented in table \ref{tab:IPv6AddressTypes}



\begin{sidewaystable}[h!]
\centering
\small
\caption{IPv6 address types and assignment }
\label{tab:IPv6AddressTypes}
\rowcolors{1}{lightgray!60}{white}
\begin{tabular}{m{1.5cm}m{2.5cm}m{14cm}m{3cm}}
\toprule
Address   Types & Address Prefix & Designation   \& Explanation & Remarks \\
\midrule
 & ::/128 & Unspecified (all zeros), used for own address identification & Equiv. to 0.0.0.0 address in IPv4 \\
 & ::1/128 & Loop back address defined for self-communication & Equiv. 127.0.0.1 address in IPv4 \\
 & ::\textless{}IPv4\textgreater{}/96 & IPv4 compatible IPv6 address, any IPv4 router having its unique IPv4 address can be identified with its compatible IPv6 address. These can be used for transition in automatic tunneling. &  \\
 & ::FFFF:\textless{}IPv4\textgreater{}/80 & IPv4 Mapped IPv6 address, used to embed IPv4 addresses into IPv6 address. It is generally used in in dual stack transition where IPv4 addresses can be mapped into an IPv6 address {[}RFC4038{]} &  \\
 & \textless{}IPv6-Prefix\textgreater{}::5EFE:\textless{}IPv4-address\textgreater{} & ISATAP   - Intra Site Automatic Tunnel Addressing Protocol consists of 64 bits global   or link local unicast IPv6 prefix. ISATAP is designed for transporting IPv6   packets within a site where a native IPv6 infrastructure is not available. &  \\
 & FE80::/10 & Link Local Unicast address. These are addresses are used for communication among hosts attached in a single link or non-routed common access network. IPv6 routers can use this address for protocol signaling like OSPF routers use this address for internal communication. &  \\
 & FC00::/7 & Unique Local Address. These are used for private communication and are not publicly routable. But are routed within the enterprise or organization. & Equiv. to IPv4 private addresses  \\
 & 2001:2::/48 & Benchmarking, these are reserved in use for documentation. & Equiv. IPv4 198.18.0.0/15  \\
 & 2002:\textless{}IPv4\textgreater{}::/48 & 6to4 address prefix. These are used by gateway router for 6to4 tunneling. &  \\
\multirow{-10}{*}{\rotatebox[origin=c]{90}{Unicast Address}} & 2000::/3 & Global unicast address. Assigned by IANA with 3 bits prefix (001) for current IPv6 address   distribution worldwide. &  \\
Multicast Address & FF00::/8 & IPv6 multicast address. The last nibble of the first block define the scope and   second last nibble specify the flag bits. Rest of 112 bits is the group identifier. Flag value of 1 specify the temporary multicast while 0 indicates   the well know multicast address. Similarly the scope value like 1-\textgreater interface local, 2-\textgreater{}Link local, 5-\textgreater Site local scope and E indicates global scope. Rest are used for future use. & Equiv. to IPv4 Class D address range. \\
Anycast Address & Any   Global Unicast IPv6 address & Anycast address shall be any one unicast address and can assign only to routers. &  \\
\bottomrule
\end{tabular}
\end{sidewaystable}

\subsection{IPv6 network migration approaches}
\label{sec:IPv6MigrationApproaches}
In this section, we discuss the major transition techniques and their implementation scenarios for service providers to migrate their existing network into IPv6 network. 

\begin{figure}
    \centering
    \includegraphics[width=\textwidth]{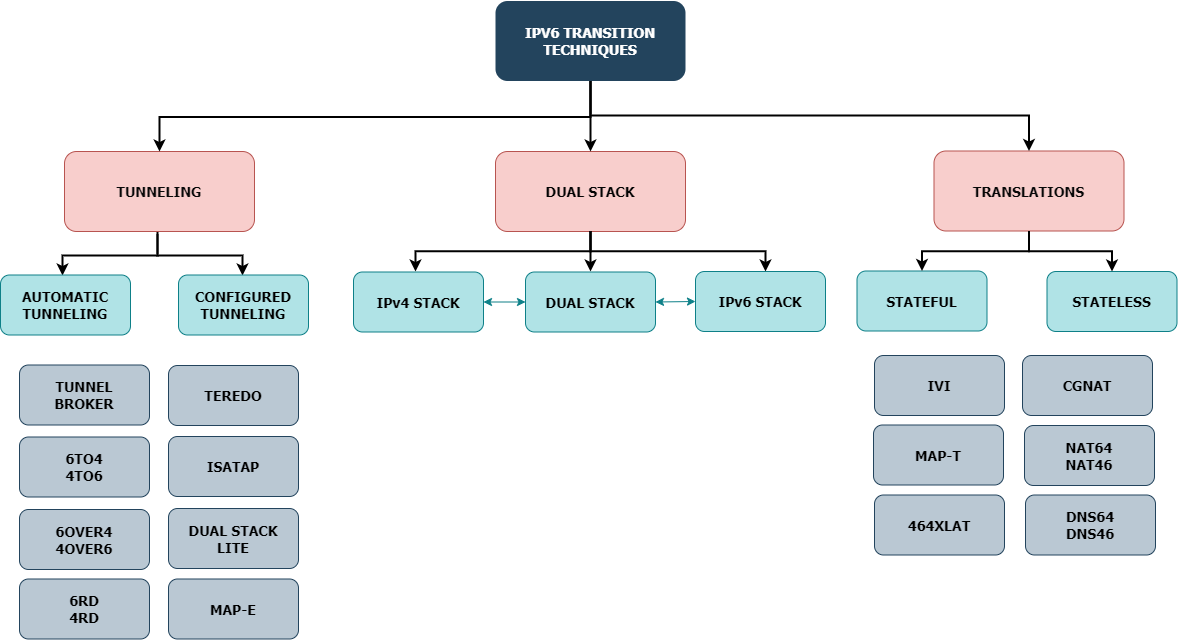}
    \caption{IPv6 network transition approaches }
    \label{fig:IPv6TransitionApproaches}
\end{figure}

IPv6 is not backward compatible. The current IPv4 protocol with application support software and hardware system needs to be upgraded to operate with IPv6. Due to varying nature of device lifetime, performances, protocols, and applications support, all the networking devices in an enterprise can’t be upgraded. In this situation, upgradability of network device needs to be identified and non-upgraded devices are to be replaced with newer hardware including protocols and application supports. Unfortunately, due to higher investment cost with respect to network size, migration at once is not possible. Hence, different transition techniques are defined and adopted in practices. The taxonomy of different transition approaches are presented in Figure \ref{fig:IPv6TransitionApproaches}. 

The deprecated NAT-PT \cite{tsirtsis2000network} in the beginning created good platform to develop more specific translation approach under stateless and stateful translation. IVI is the mechanism that performs IPv6 header, transport layer header, and ICMPv6 header translation and vice versa. Stateful IVI consists of 1:1 and 1:N translation. 1:1 means one IPv4 address is mapped to one IPv6 address while 1:N translation has one IPv4 address is translated into many IPv6 addresses. To achieve better throughput, stateless IVI is applied in the backbone network \cite{Zhai2011}. Stateful IVI is used to connect IPv4 hosts to IPv6 internet and are used mostly at edge network because of its higher computational cost. 

NAT64 is a mechanism that provides stateless approach to translate IPv4 header into IPv6 and vice versa. IPv4 address is directly embedded into the IPv6 address. Major drawback of NAT64 stateless translation is that it only translate IPv4 options which has IPv6 counterparts. No state/bindings created on the translation (statelessness). It is 1:1 translation. It does not translate IPv6 extension headers beyond fragmentation extension header. It assures end-to-end transparency. It allocate IPv4 address for each IPv6 only device that requires translation. It does not solve the problem of IPv4 address depletion. For one IPv4 to many IPv6 addresses mapping, stateful NAT64 is required. It is a 1:N translation. It multiplexes many IPv6 devices into a single IPv4 address. Suitable to apply when IPv6 only device want to communicate with IPv4 internet. State/bindings are created on every unique translation. It only supports IPv6 initiated flows. DNS64 is required for both stateful and stateless NAT64. It does not assure end-to-end transparency. NAT46 allows an IPv4-only client to communicate with an IPv6-only server by translating IPv4 header to IPv6 and vice-versa for the return traffic. It needs DNS46 for large network to work. In general NAT464 are called address family translator (AFT). Many mobile networks may need this approach to have battery saving of the mobile client. ISPs may choose this inside the residential gateways for them to connect to corresponding other network. 

464XLAT \cite{mawatari2013464xlat} is the combination of stateful and stateless translation. CLAT is the customer-side translator of XLAT. It implements RFC6145 as stateless NAT46 approach while PLAT is the provider-side translator of XLAT that implements RFC6146 as stateful NAT64.

Carrier grade NAT (CGNAT) is the marketing term also called the large scale NAT that implements the translation between IPv4 private to IPv4 public address called NAT444, which is in fact a recursive NAT. It also has the multiple functionality of including NAT64 and hence CGNAT comes in different flavor like NAT444, NAT464 including DNS64. MAP-T is the implementation of double stateless NAT64 based solution for providing shared or non-shared IPv4 address connectivity \cite{bao2013ivi}. 

\subsection{Worldwide IPv6 Network Deployment Status}
\label{sec:SDNSection}
Almost 28\% of devices on the Internet are currently IPv6 capable \cite{APNIC}, while the capability by continent in America, Asia, Europe, and Africa are 33\%, 29\%, 21\% and 1.2\% respectively. 

Different IPv6 transition methods discussed in Section \ref{sec:IPv6MigrationApproaches} are not mutually inclusive. "\textit{Which method is suitable to implement for transition?}" is generally depends upon the service provider’s network status and their sole decision. For example, if an ISP is in early stage of its IPv6 network migration, but there is customer demand of IPv6 based services, then 6RD technique for quick service delivery would be suitable. Similarly, if ISP backbone network is already IPv6 ready, then DS lite or address family translation e.g. XLAT is suitable. Looking into the world’s largest service providers, Google has already migrated its enterprise network into IPv6 \cite{babiker2011deploying}. Similarly, AT\&T, NTT, and other largest telecom operators have their network operation successfully running with IPv6. In most of the countries, IPv6 network migration is guided by the national regulatory policies \cite{Tadayoni2016} that ISPs are migrating their network to IPv6 operable network accordingly.


\section{Networking with SDN}
\label{sec:smartNetwithSDN}
\subsection{Introduction and Definition}
The World-wide network infrastructure is growing with respect to the service demands. The currently operating network is highly heterogeneous, in which there is required to deal with different protocols, many platforms, and vendor specific network equipment with their own proprietary software. Network operators have to configure their individual network devices separately using either low-level or vendor specific configuration commands. This individual device configuration is highly time consuming and complex in network management, maintenance, and troubleshooting. The control plane takes decision to handle network traffic and directs data plane devices based on its decision. Data plane forwards traffic according to the decisions made by the control plane. The data and control plane functionalities bundled inside the single networking device reduces flexibility, hindering innovation, and evolution of the networking infrastructure \cite{kreutz2014software}. Flexibility in the networking operation and management with many other features via programming is introduced by the concept of SDN and NFV. It was originally coined with the ideas by conceptualizing "OpenFlow" at Stanford University, USA. The definition of SDN is:

\textit{"Software-defined networking (SDN) is an emerging network paradigm where network control plane is decoupled from forwarding plane and is directly programmable”} \cite{nadeau2013sdn}.

\begin{figure}[ht!]
    \centering
    \includegraphics[width=0.8\textwidth]{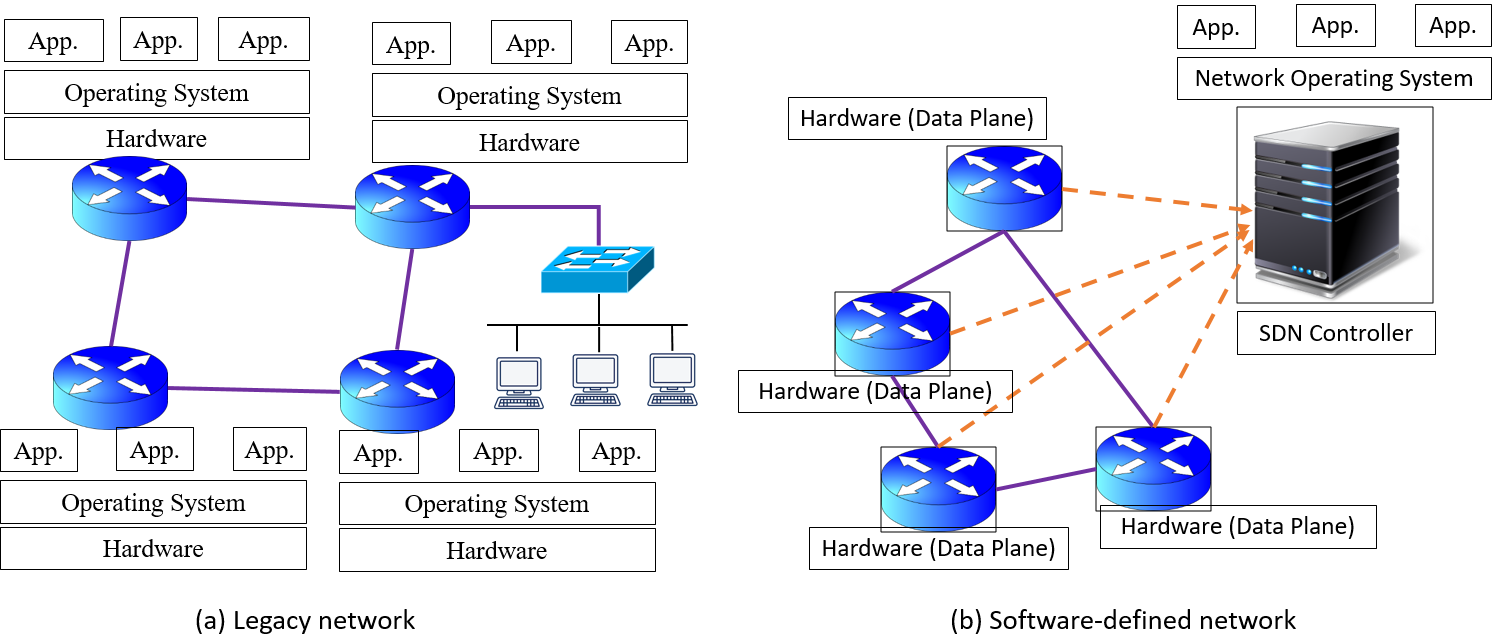}
    \caption{Functional architecture of (a) legacy network, and (b) the SDN}
    \label{F2:ArchSDNLegacy}
\end{figure}

The forwarding state in the data plane is remotely controlled/managed by the decoupled control plane called SDN controller, which is located as an external entity, for example, at the network operation center (NOC). A simple network consists of legacy networking devices is shown in Figure \ref{F2:ArchSDNLegacy}a \cite{dawadi2019software}. Legacy networking device has its control and data plane vertically integrated that makes the network complex in management and configurations. Individual legacy device also consists of application layer services that includes routing, mobility, access control, virtual private networks (VPNs), multi-protocol level switching (MPLS), traffic, and security management etc. Figure \ref{F2:ArchSDNLegacy}b presents a simple network architecture, where the control plane and application layer services are detached from individual network devices and moved to a logically centralized controller. The control plane and north bound applications run on a commodity hardware called an SDN controller, whose major task is to control the communications in data plane devices.

The major driving factors of SDN are briefly summarized below.

\begin{enumerate}[label=\roman*.]
    \item \textbf{Separation of control and data plane}\\
    Individual device control planes are removed and centralized into the SDN controller. The network device's data plane merely works as a packet forwarding engine based on the controller's decision. This logical centralization of the controller opens up a lot of possibilities for developing bespoke apps and implementing network policies through abstraction on the northbound side. This makes the network more flexible in terms of operation and control by reducing the complexity of networking functions, applications, and network services.
    
    \item \textbf{Flow based} \\
   The SDN controller makes forwarding decisions based on flow rather than destination \cite{kreutz2014software}. In SDN, a flow is defined as a set of packet field values that serve as a match (filter) criterion. It is made up of a series of operations (instructions) performed on the packet sequence from source to destination.
    
    \item \textbf{Programmable network}
    SDN's most crucial characteristic is its programming ability. It's extremely flexible, allowing custom software applications built on top of the SDN controller to connect with data plane devices for network management and administration.
    
    \item \textbf{Open interface}\\
    The networking system can be used as a vendor-neutral common platform for network administration by standardizing an open interface with open application programming interfaces (APIs) and communication protocols, such as OpenFlow, between devices with control plane (SDN controller) and data plane.
    
    \item \textbf{Abstraction}\\
    To accommodate equipment from a number of vendors and technologies, as well as to enable the control plane to service a variety of applications, SDN applications are isolated from their underlying network technologies.
    
    \item \textbf{Security}\\
    With the centralization of the control plane in the network, network programmability increases the freedom to apply diverse security policies to create a robust and highly secure network environment in SDN.
    
    \item \textbf{Energy efficiency}\\
    Energy consumption by network equipment is higher in the legacy IPv4 networking system due to the lack of smart controlling mechanisms. As the size of the network grows, so does the energy bill. SDN is more energy efficient, with energy savings gained either through algorithmic or hardware enhancements \cite{dawadi2020towards}. Implementing SDN has significant OpEX saving with energy optimization and reduction of CO2 emission making the network more energy aware to promote green ICT \cite{dawadi2020towards,Dawadi2019a}.
\end{enumerate}

\begin{figure}[h]
    \centering
    \includegraphics[width=0.6\textwidth]{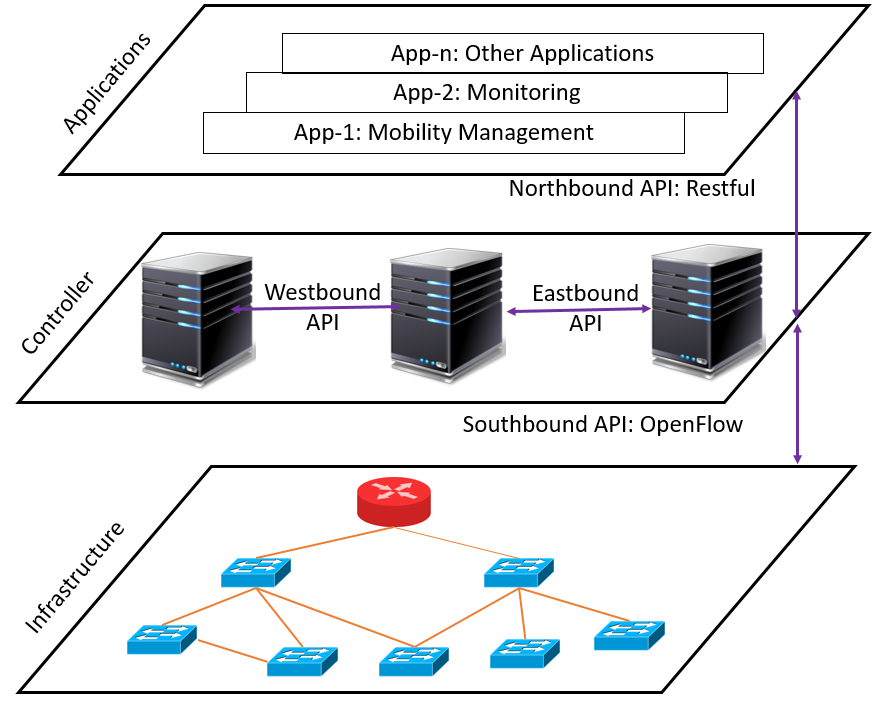}
    \caption{SDN layered architecture}
    \label{F2:SDNLayerArch}
\end{figure}

The structure of SDN is divided into three layers viz. infrastructure layer (data plane) in the bottom, controller layer at the middle, and application layer in the top. The bridging between infrastructure and applications in SDN is done by controller through its middleware called northbound and southbound APIs. For the controller load balancing with reliability, efficiency, and fault tolerant, additional controllers can be attached via eastbound and westbound APIs. Northbound APIs are RESTful APIs like frenetic, xml, json etc., while OpenFlow is a popular vendor neutral protocol available in southbound to establish communications between the controller and the data plane devices. Figure \ref{F2:SDNLayerArch} depicted the overall layered architecture of SDN \cite{dawadi2019software}.

Principally, SDN capable switch checks the flow entries of packets incoming and directs the packets to outgoing interface, if the flow entry is matched. Otherwise, the packet header, whose flow entry is not found in the flow table, would be forwarded to the controller for further decision. The controller provide decision by updating flow tables of the entire switches in its controlled network towards the path of the packet's destination.

\subsection{Need and Benefits of SDN}
\begin{figure}
    \centering
    \includegraphics[width=\textwidth]{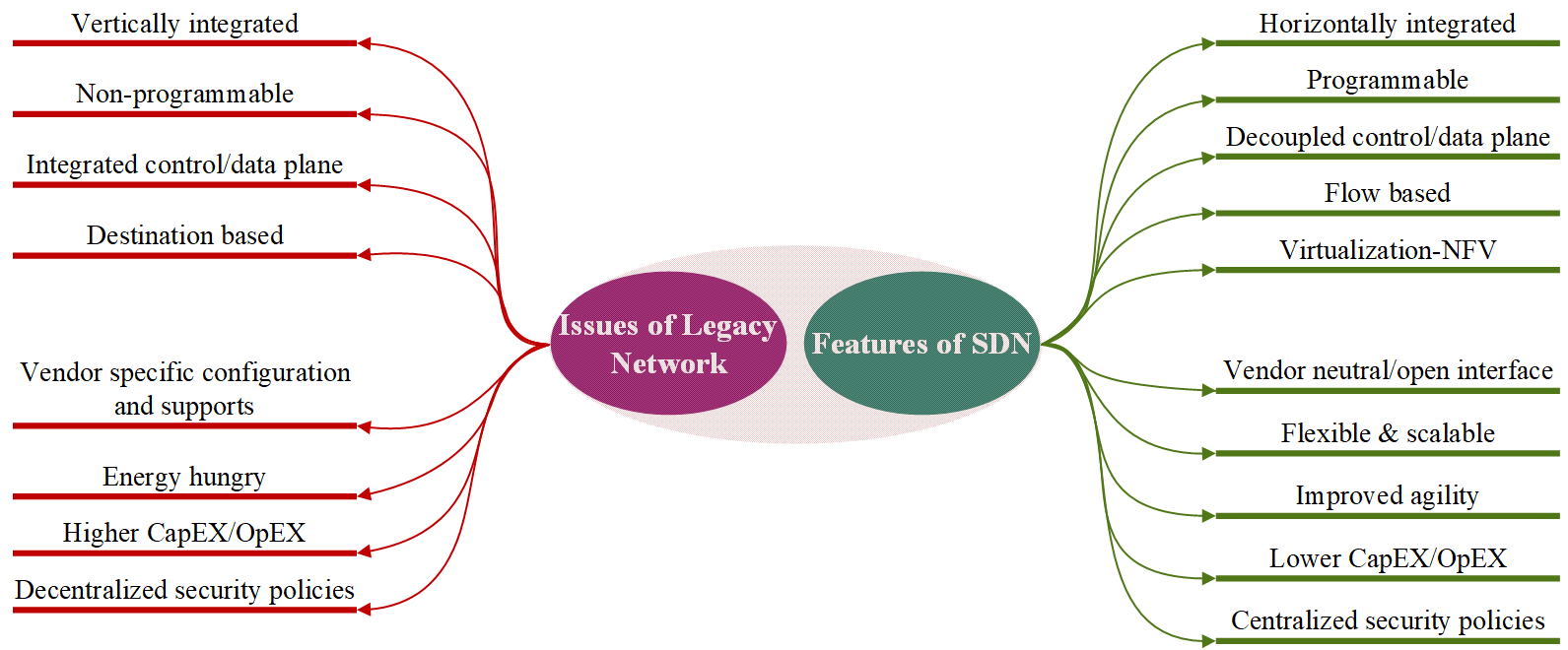}
    \caption{Problems of legacy network and features of SDN}
    \label{F2:LegacySDNIssueFeature}
\end{figure}

SDN allows network operators to manage and operate virtualized resources without deploying additional hardware. This paradigm shift in network operation and management is considered as the advanced networking approach that counters the increasing complexities in the existing legacy networking system and optimize the operational cost. The existing problems of legacy IPv4 networking system and the features of SDN are listed in Figure \ref{F2:LegacySDNIssueFeature}.

SDN increases automation in network management and operation with less human intervention that could help to reduce the CapEX and OpEX of the organizations \cite{rizvi2015towards,Hernandez-Valencia2015,karakus2018economic}. Hence, it encourages the service providers to search for the better options and attraction towards SDN. Besides the implementation challenges \cite{sezer2013we}, SDN is proven technology towards efficient network management that it solves those existing issues in the legacy IPv4 network and create highly flexible, visible, programmable, scalable, modular, open interface, and abstraction-based networks \cite{brief2014sdn,raza2014comparison}. Migration to SDN over data center networks are popularly endorsed \cite{Goransson2014,dai2017enabling,AT&T2014}, while ISP networks migration is in the early stages. Similarly, the ongoing research, development, implementation, testing, and verification   \cite{brief2014sdn,csikor2018harmless,AT&T2014,kobayashi2014maturing,ON.LAB2014,babiker2011deploying} of SDN and IPv6 implementations in ISP/Telcos network are encouraging activities for service providers to migrate their legacy networks in a phase-wise manner.

\subsection{SDN Migration Approaches}
\label{sec:SDNMigrationApproaches}
The changes in the networking paradigm by detaching control plane form each switch/router and integration into a single controller to manage/control entire network through remote controller is the major technology change endorsed by SDN. This paradigm shift has several benefits as compared with legacy networking. Significant amount of OpEX and CapEX saving can be achieved with the implementation of SDN \cite{Hernandez-Valencia2015,bogineni2014introducing}. But the challenging situation with SDN migration is the issue of network device upgrades or replacement as immediate migration to SDN is not viable same as that of IPv6 network. Smooth transition approaches for SDN migration in ISP and Telcos networks are still an ongoing research and implementation stage. Three approaches of legacy network migration to SDN proposed by ON.LAB \cite{brief2014sdn} are depicted in Figure \ref{F2:SDNMigrationApproach}.

\begin{figure}[h]
    \centering
    \includegraphics[width=0.8\textwidth]{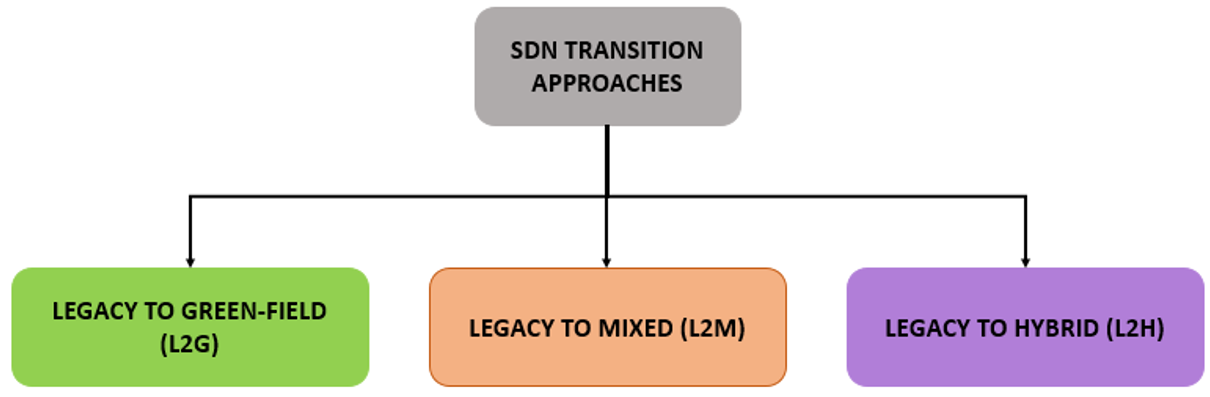}
    \caption{SDN migration approaches}
    \label{F2:SDNMigrationApproach}
\end{figure}

The legacy to greenfield (L2G) approach enables clean setup of the SDN in which either only the complete replacement of existing networking infrastructure or expansion of new network is possible. But for the running network infrastructure, the immediate replacement of network devices is not viable for service providers due to many complexities, e.g. higher cost of investments, lack of skilled HR etc. Hence, this approach is applicable for service providers to expand purely new network with possible pilot tests and experimentation.

The legacy to mixed (L2M) approach supports a gradual transition to SDN, while the network consists of legacy and SDN enabled devices during migration. Once the interoperability between legacy devices and SDN devices is ensured, this approach is fairly suitable for migration implementation. SDN-IP implementation over ONOS enables integration of legacy networks with SDN for the mixed types of communication in a multi-domain routing environment \cite{Dawadi2020, bogineni2014introducing,7502479}.

Routers in the legacy to hybrid (L2H) approach maintains both legacy routing and OpenFlow table. The router in this approach is supposed to be a dual-stack device having the option to forward traffic based on legacy routing or OpenFlow table based on the incoming traffic to be destined. Experiment by ON.LAB \cite{brief2014sdn} and studies by different authors \cite{levin2014panopticon,vissicchio2015central,csikor2018harmless} indicated that migration to hybrid network is viable. These days, the terms 'mixed' and 'hybrid' are used interchangeably \cite{amin2018hybrid,vissicchio2017safe}. Hybrid networks or mixed networks both constitute the existence of SDN and non-SDN devices in the same network. Hybrid switch means it is capable to operate legacy routing and OpenFlow forwarding both. In summary, legacy network migration to SDN using L2G approach is not a viable solution, while a complete set of new infrastructure can be established using this approach. From the migration perspectives, L2H/L2M approaches are more reasonable to follow for smooth transition.

Being an ongoing research work, development of approaches for SDN migration in Telcos/ISP networks domain have higher priority to the World-wide researchers. There are some contributory works regarding the implementation of L2H/L2M approach that attempted to provide the paths for real time transition of legacy networks into SDN. For example, HARMLESS \cite{csikor2020transition}, OSHI \cite{salsano2015hybrid}, Panopticon \cite{levin2014panopticon}, RouteFlow \cite{vidal122013building}, and Fibbing \cite{vissicchio2015central} are some of the approaches proposed for the migration to hybrid IP/SDN. However, the implementation of any approach proposed in the production network is not known. Recently, ONOS/SDN-IP \cite{ON.LAB} is the dedicated controller and application developed by ON.LAB for carrier grade network migration of ISP/Telcos networks into SDN. Figure \ref{fig:LegacyMixedHybridNetwork} shows the glimpse of legacy and mixed/hybrid networks.

\begin{figure}[ht!]
   \centering
   \begin{subfigure}[b]{0.30\textwidth}
         \includegraphics[width=\textwidth]{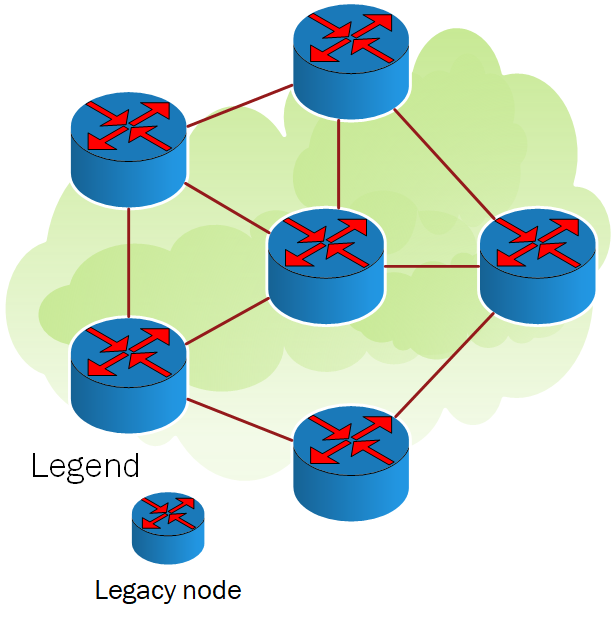}
         \caption{Legacy network}
         \label{sb:LegacyNetwork}
     \end{subfigure}
     \hfill
    \begin{subfigure}[b]{0.30\textwidth}
         \includegraphics[width=\textwidth]{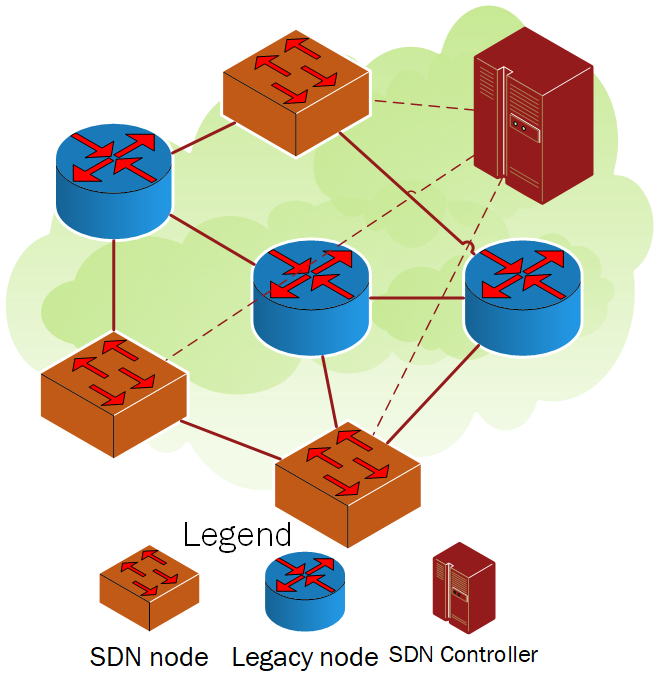}
         \caption{Mixed/hybrid network}
         \label{sb:MixedNetwork}
     \end{subfigure} 
    \hfill
     \begin{subfigure}[b]{0.30\textwidth} \ContinuedFloat
         \includegraphics[width=\textwidth]{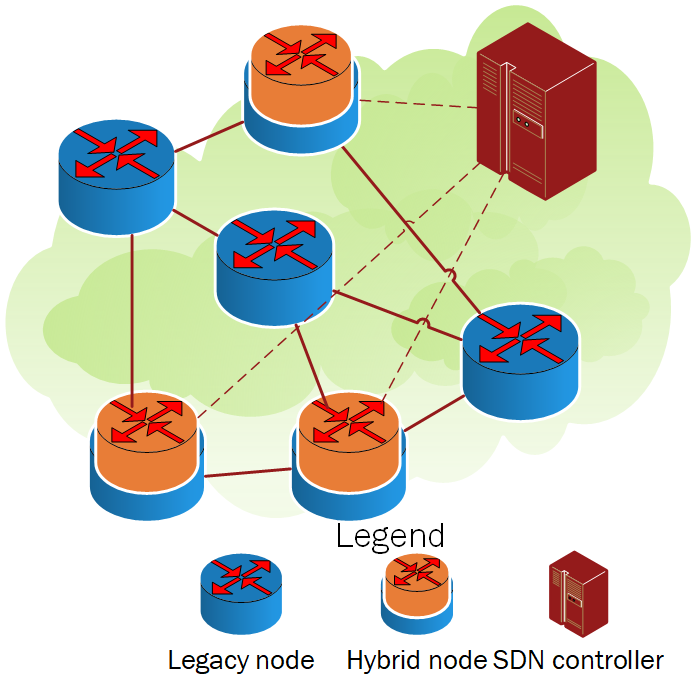}
         \caption{Mixed/hybrid network}
         \label{sb:HybridNetwork}
     \end{subfigure}
    \caption{Legacy, mixed, and  hybrid network scenarios}
    \label{fig:LegacyMixedHybridNetwork}
\end{figure}

\begin{figure}[ht!]
    \centering
    \includegraphics[width=0.8\textwidth]{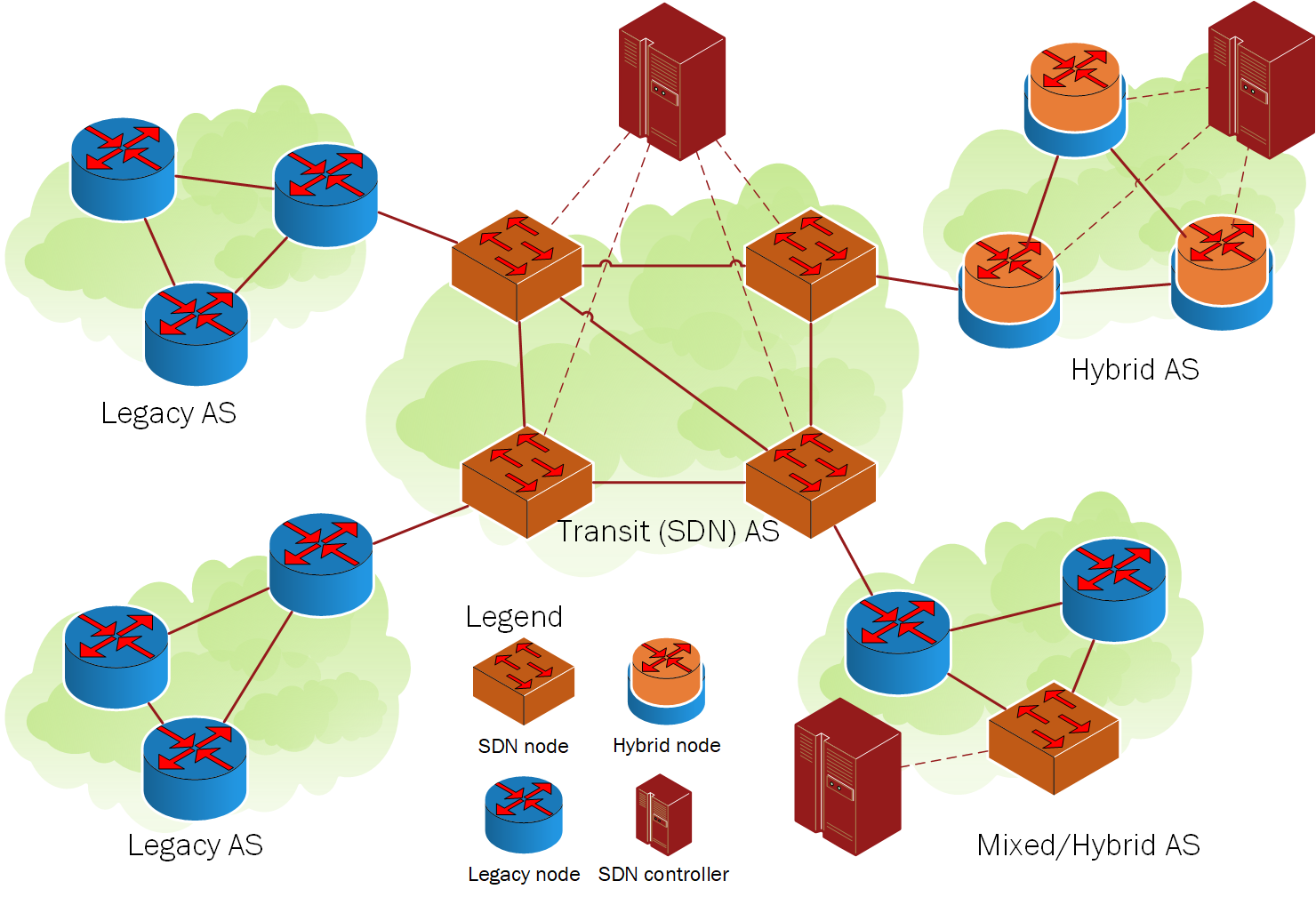}
    \caption{ISP hybrid SDN. AS, autonomous system}
    \label{fig:ISPHybridNetwork}
\end{figure}

Figure \ref{fig:ISPHybridNetwork} presents the use case of multi-domain ISP network constituting use of legacy, SDN, and hybrid switches to show a mixed/hybrid ISP network to be realized to establish inter-AS communication between SDN and legacy network using ONOS/SDN-IP \cite{Dawadi2020Legacy}. Figure \ref{sb:LegacyNetwork} shows the legacy network structure with legacy node (router). During the network migration, some switches can be fully SDN compatible while some still remain as legacy switches as shown in Figure \ref{sb:MixedNetwork}. Similarly, if a switch maintain legacy routing table as well as OpenFlow table both and forward the traffic based on the traffic incoming to it, this is a hybrid switch that runs in dual stack mode (legacy and SDN) as shown in Figure \ref{sb:HybridNetwork}.

\section{Joint SDN and IPv6 Network Migration}
\label{sec:jointSDNIPv6}
In the previous sections, we discussed separately about the SDN and IPv6 network. Both are networking paradigms explicitly require for ISPs to be considered for the migration. Hence, we introduced joint SDN and IPv6 network as the "Software-Defined IPv6 (SoDIP6) Network" for implementation and migration consideration. 

SoDIP6 network is defined as: \\
\textit{"The complete network and server systems operated with IPv6 addressing and routing over Software-defined network environment, in which the data plane forwarding devices enabled with IPv6 packet communications, are controlled and managed by the logically centralized SDN controller."}

\begin{figure}[ht!]
    \centering
    \includegraphics[width=0.8\textwidth]{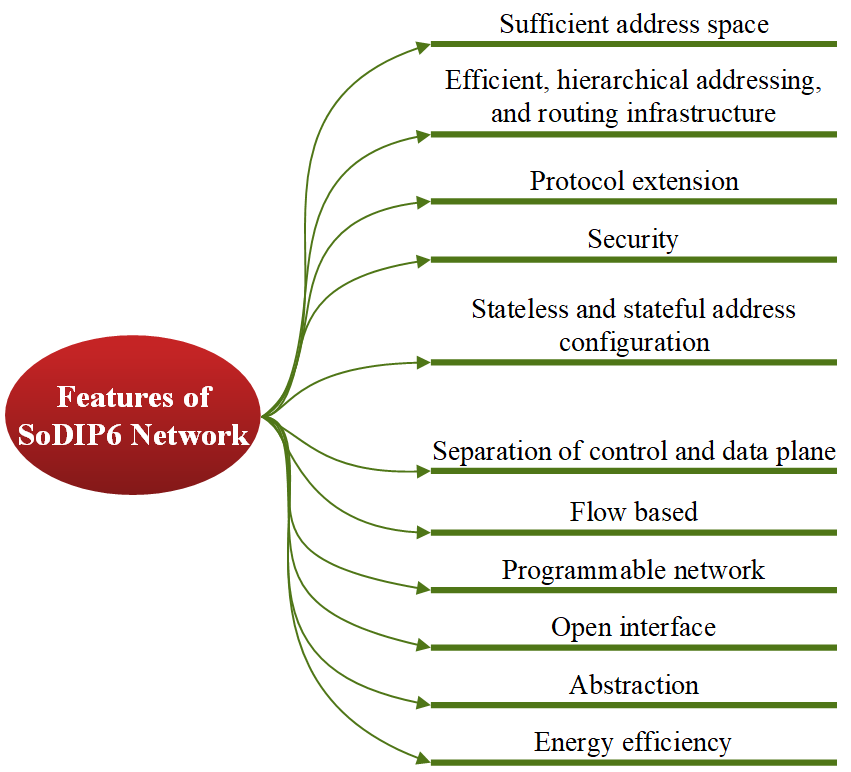}
    \caption{Features of SoDIP6 network}
    \label{F3:SoDIP6Features}
\end{figure}

Considering joint network migration as hybrid SoDIP6 network, in which the legacy IPv4 networks co-exists with the SoDIP6 network during the period of transition so that overall network migration cost can be minimized. Details about the joint network migration in terms of CapEX and OpEX optimization can be found at \cite{Dawadi2020Migration}. Similarly intelligent approaches to joint network migrations are available at \cite{dawadi2019evolutionary, Dawadi2020Legacy}. The overall features of SoDIP6 network are the combined features of SDN and IPv6. All present network system concerns, such as address depletion, NAT proliferation, vendor-specific setup, control, and operation complications, may be avoided solely by implementing the SoDIP6 network. Although ISPs can continue to use the legacy IPv4 system for a longer time, translations and tunneling options are becoming more expensive and complex to operate and administer with the growing network infrastructure and the Internet users. The features of SoDIP6 network that encourages for network migration are shown in Figure \ref{F3:SoDIP6Features}.

\begin{enumerate}[label=\roman*.]
    \item \textbf{Sufficient address space}\\
    One hundred and twenty eight bit length IPv6 addressing structure using hexadecimal numbers provides higher than the astronomical value to uniquely identify networking devices in this universe. This can create flexible and scalable network, while the implementation of IoT and expansion of WSN will be more convenient to create smart World by using IPv6 addressing.
    
    \item \textbf{Efficient addressing and routing infrastructure}\\
    Internet assigned numbers authority (IANA) has defined the hierarchical distribution of global IPv6 addresses starting with global routing prefix then to regional internet registries, national internet registries, and local ISPs \cite{iab2001iab,narten2011ipv6}. This creates an efficient, hierarchical, and summarized routing infrastructure.

    \item \textbf{Stateless and stateful address auto-configuration (SLAAC)}\\
    Address auto-configuration is a new IPv6 addressing capability that allows both stateless and stateful addressing to configure host addresses automatically. In SLAAC, an IPv6 host automatically configures its link local and global IPv6 addresses using a random assignment technique or the EUI-64/SEUI-64 address format \cite{Abdullah2019} to specify the IPv6 suffix, whilst prefixes are promoted by local routers. IPv6 provides the same notion as IPv4 in terms of stateful addressing via DHCP.
    
    \item \textbf{Protocol extension}\\
    IPv4 headers are limited to 40 bytes of optional header fields, whereas IPv6 headers can readily take extensions with new features, which are maintained in a daisy chain fashion after the IPv6 main header.
    
    \item \textbf{Separation of control and data plane}\\
    Individual device control planes are removed and centralized into the SDN controller. The network device's data plane simply operates as a packet forwarding element based on the controller's choice. This logical centralization of the controller allows for the development of bespoke applications and the implementation of network policies via abstraction at its northbound. This makes the network more flexible in terms of control, operation, and management by reducing the complexity of networking functions, applications, and network services.
    
    \item \textbf{Flow based} \\
    The SDN controller makes forwarding decisions based on flow rather than destination \cite{kreutz2014software}. In SDN, a flow is defined as a set of packet field values that serve as a match (filter) criterion. It is made up of a series of operations (instructions) performed on the packet sequence from source to destination.
    
    \item \textbf{Programmable network}\\
    The capacity to program is the most important feature of SDN. It is highly customizable, allowing bespoke software applications built on top of the SDN controller to communicate with data plane devices for network administration and management.
    
    \item \textbf{Open interface}\\
    The networking system can be used as a vendor-neutral common platform for network administration by standardizing an open interface with open APIs and communication protocols like OpenFlow between devices with control plane (SDN controller) and data plane.
    
    \item \textbf{Abstraction}\\
    SDN applications are abstracted from their underlying network technologies to support equipment from many vendors and technologies, as well as to enable the control plane to serve a variety of applications.
    
    \item \textbf{Security}\\
    IP Security (IPSec) is a default security framework defined under IPv6 protocol suite requirement. IPSec provides set of standards for authentication and encapsulation with key management framework for network security needs and promotes interoperability between different IPv6 implementations. Similarly, network programmability with centralization of control plane in the network adds more flexibility to apply different security policies to build robust and highly secure network environment.
    
    \item \textbf{Energy efficiency}\\
    Due to the lack of smart controlling features in the legacy IPv4 networking system, energy consumption by network equipment is higher. The energy bill increases with increasing network size as well. SoDIP6 network is more energy efficient, in which energy saving can be achieved algorithmically or through the hardware improvements \cite{dawadi2020towards}. Implementing SoDIP6 network has significant OpEX saving with energy optimization and reduction of CO2 emission making the network more energy aware to promote green ICT \cite{dawadi2020towards,Dawadi2019a}.
\end{enumerate}

\subsection{Legacy Network Migration to SDN and IPv6}
\label{NetMigration-Challenges}
\subsubsection{\emph{Network migration challenges}}
\begin{figure}
    \centering
    \includegraphics[width=\textwidth]{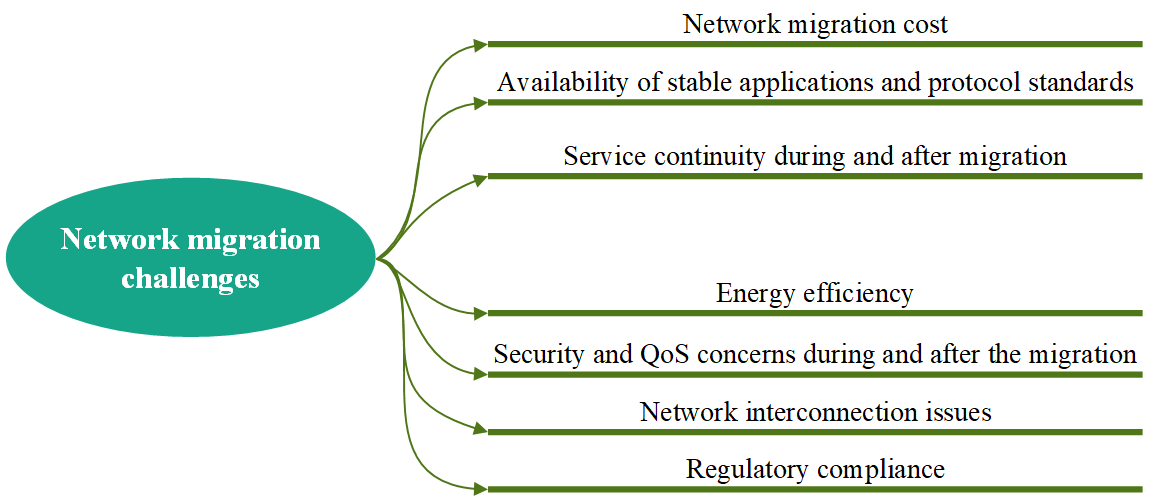}
    \caption{Network migration challenges.}
    \label{F2:NetMigrationChallenges}
\end{figure}

Network migration is a complex process because the networking infrastructure can’t be transformed on-the-fly leading to a delay in the migration \cite{Yadav2012} due to major challenges \cite{rizvi2015towards,Dawadi2019,Levin2014} highlighted in Figure \ref{F2:NetMigrationChallenges}. The network migration challenges for service providers are briefly discussed here.

\begin{enumerate}[label=\roman*.]
\item \textbf{Network migration cost}\\
The issues of investment and operation cost have many folds. Network infrastructure consists of large number of networking components that are not possible to upgrade or replace on-the-fly. Enterprises should be financially ready, human resource (HR) ready, application and protocols standards ready, and ready for disaster management as well \cite{Main2015}. For the fairly sustained service providers of developing nations like Nepal, it is hard to migrate their infrastructure into operable latest technologies like SDN and IPv6.
    
\item \textbf{Availability of stable applications and protocol standards}\\ 
Readiness measurement on network and server applications as well as protocol standards and their applicability are important before planning for migration. The World-wide deployment progress of IPv6 shows that the applications and protocol standards are well tested and implemented, while sufficient benchmarking on SDN related network applications and protocols are still an ongoing process. It is required to evaluate all the networking components e.g. hardware, operating system (OS), applications, security systems, and many more for the better continuity of services.
    
\item \textbf{Service continuity during and after the migration}\\
The changes on the system may interrupt the overall networking operations. This is one of the major factor that delays the migration. Service providers are always in fear of service interruption, while approaching the new technology in the implementation. 

\item \textbf{Energy efficiency}\\
The expansion of network size including hardware and software increases energy bill annually \cite{dawadi2020towards}. The optimization of network equipment energy usage with an emphasis on green computing in SoDIP6 network operation is also a major concern for service providers. 

\item \textbf{Concerns of security and QoS during the migration}
Another key problem is ensuring that the network is secure during and after migration. It goes without saying that implementing new technologies and integrating them with old systems throughout the transition can pose a security risk. The security of firewalls, servers, and other applications should also be able to keep up with the latest technologies.

\item \textbf{Network interconnection issues}\\
The World-wide network/internet architecture is a hierarchical tired architecture. Tier-1 ISPs are core network/internet service providers, Tier-2 ISPs are transit service providers, while Tier-3 ISPs are the service providers of end-access customers. The sources of contents in the internet are highly distributed and heterogeneous. ISPs are fully interconnected. They have their own trade agreement and management to exchange the traffic. In this aspect, only migrating of one ISP network does not have meaning and could not provide the latest services to its customer. It is required to have a close coordination with other interconnected ISPs in the same level and in the hierarchy.
    
\item \textbf{Regulatory compliance}\\
International regulatory bodies like international telecommunications union (ITU) and Internet corporation for assigned name and numbers (ICANN) have issued standards procedure to IPv6 network migration. Similarly, every country have their own standards and regulatory guideline for network migration \cite{Committee2012}. The uneven distribution of IPv4 addresses by the internet registries has created the imbalance on network migration activities World-wide. Asia is the region, where the public IPv4 address exhaustion was announced first time in 2011. ARIN and AFRINIC had lots of IPv4 address block till 2019 leading to delay in exhaustion \cite{jia2019tracking,ITU2020}. The regulatory standards of SDN migration is an ongoing process \cite{ITU2018,ITU2020}. Hence, this is an additional challenges for service providers to remain in the boundary of regulatory compliance to migrate their networks.
\end{enumerate}

\subsubsection{\emph{Network migration steps for service providers}}
Most of the approaches for the transition to IPv6 and SDN discussed in Sections \ref{sec:IPv6MigrationApproaches} and \ref{sec:SDNMigrationApproaches} are being adapted by different organizations World-wide \cite{kobayashi2014maturing,ON.LAB2014,lencse2019comprehensive}. The implementation of transition mechanisms depend on the running status of ISPs and their interconnection with other ISPs. In the case of new network deployment, L2G approach of SDN migration is preferable, but the existing network migration is only viable either by upgrading the running network devices or replacing it with new devices that are SoDIP6 capable. After investigating through different transition approaches for SDN and IPv6 both, it is considered that dual-stack IPv6 and hybrid SDN \cite{kobayashi2014maturing} are the best choice in the joint migration modeling for smooth transition to SoDIP6 network. The joint network migration planning is based on the World-wide tired ISP network interconnection architecture \cite{dawadi2019evolutionary}.

\begin{figure}[ht!]
    \centering
    \includegraphics[width=\textwidth]{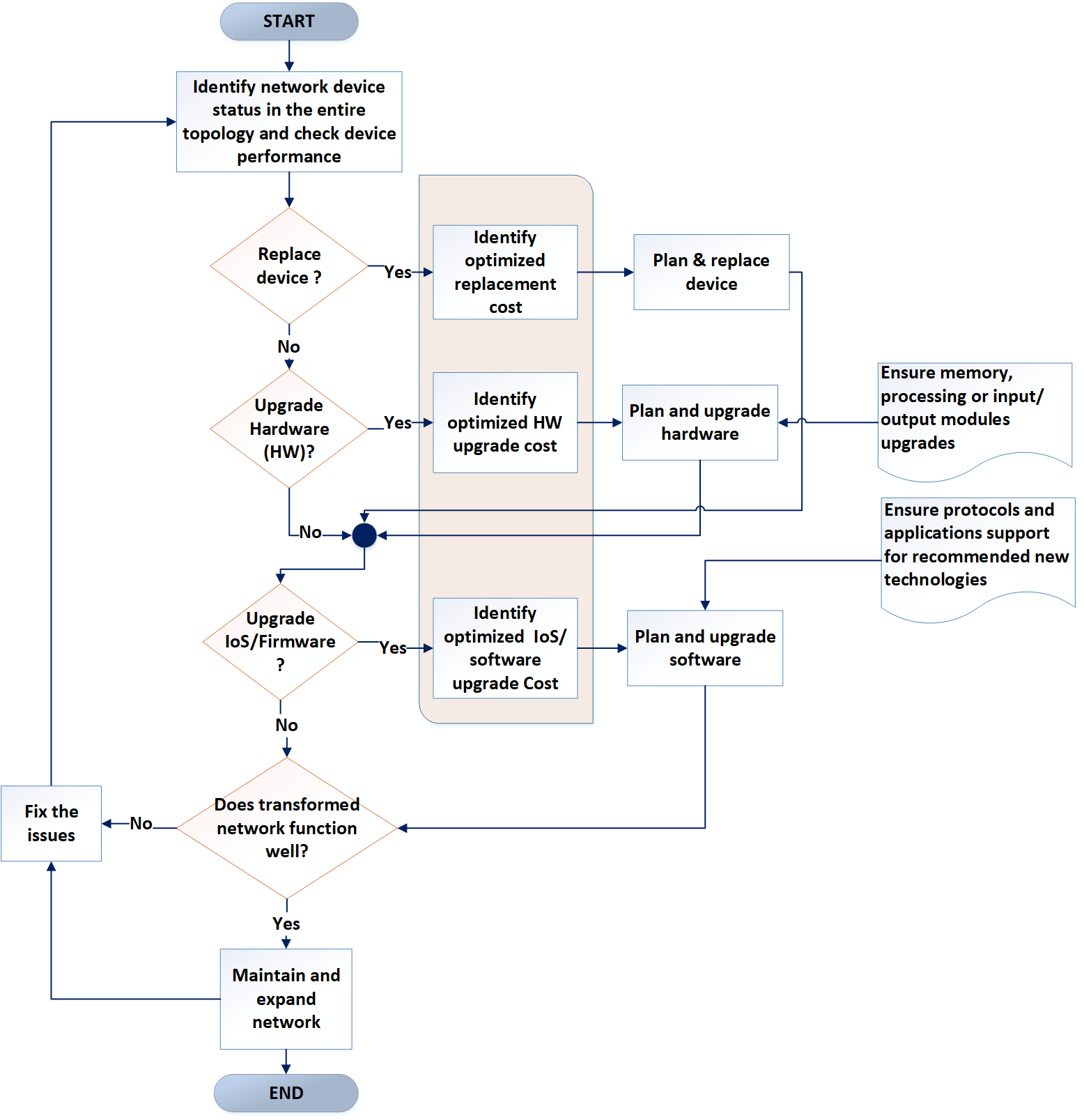}
    \caption{Network migration steps for service providers \cite{Dawadi2020Migration}}
    \label{F3:SoDIP6migrationSteps}
\end{figure}

Every service provider must keep up to date information on network devices in order to get full understanding of existing network devices and plan for proper migrations. The inventory of hardware and software details can be used to determine whether a network device can be upgraded or should be replaced in order to support newer technologies. The primary processes in network migration include device identification, budget estimation, upgrade or replacement plan development, and plan implementation. It is necessary for service providers to keep track of their network devices and infrastructure in order to monitor their status using appropriate management tools \cite{Omantek,Net.InventoryAdvisor,OCS-Inventory}. The overall steps for network migration planning are shown in Figure \ref{F3:SoDIP6migrationSteps}.

Service providers first identify the device status, whether the running devices are to be replaced or its firmware/hardware upgrade is sufficient for migration. On the status identification, total number of devices to be replaced or upgraded will be identified, then assessment of human resources with total cost of network migration will be performed. Hardware, upgrades generally means increase of memory and processing capacity of the device. In software/firmware upgrade, it is required to ensure the supports of IPv6 routing and forwarding capability, OpenFlow communication, security, and quality of service policy as well as applications and protocol supports by the upgraded device. Migration implementation phase continuously tests and evaluates the functional operation of network, when if successful migration is completed. The same steps from the beginning will be repeated with the network expansion and next phase network migration to other newer technologies.

\section{SoDIP6 Network as a Backbone of 5G}
\label{sec:5GSoDIP6}
Migration from 4G to 5G is not in the scope of this paper. However, the 5G backbone network is considered to be fully SDN based. We consider inter-relationship among IPv6, SDN, and the 5G network in this paper and discuss about the need of SoDIP6 network for successful implementation of 5G in service provider networks.  Details about migration to 5G networks considering steps/road-map from core network to transport network to end access network including proper spectrum utilization can be found at \cite{zakeri2020e2e,agiwal2021survey,suthar2020migration}.

\subsection{Introduction to 5G}
\begin{figure}[ht!]
    \centering
    \includegraphics[width=0.8\textwidth]{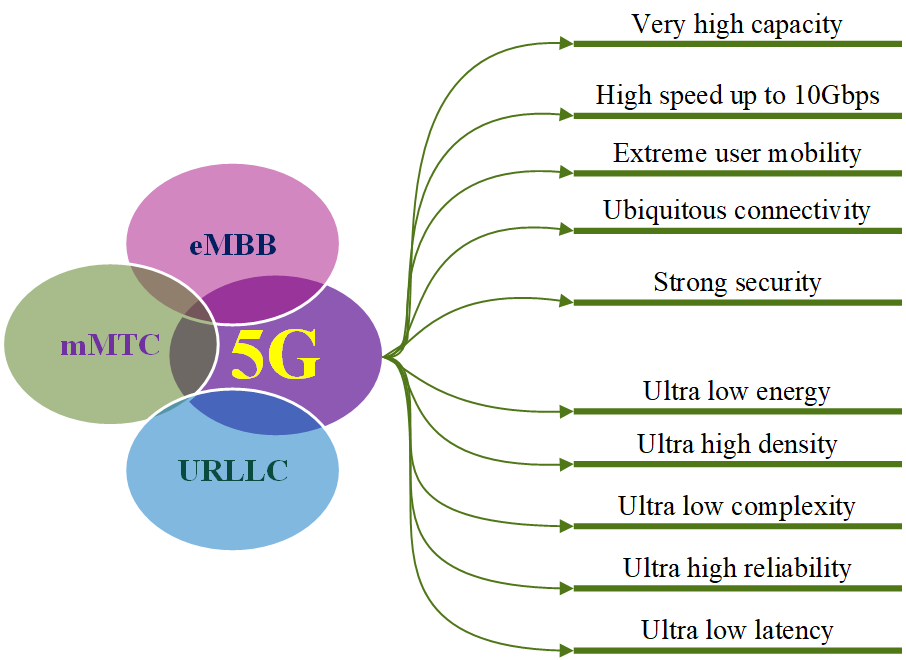}
    \caption{5G network features}
    \label{Fig:5Gfeatures}
\end{figure}

Fifth generation (5G) network is the next generation wireless networking technologies evolved with the concepts defined in third generation partnership project (3GPP) at 2015 \cite{lee2016lte}. The evolvement of 5G brings superior features and enhancement in the previous generation network and mobile communication technologies. It has major three driving features that are (i) enhanced mobile broadband (eMBB), (ii) massive machine type communications (mMTC), and (iii) ultra reliable low latency communications (uRLLC). The major features of 5G network is depicted in Figure \ref{Fig:5Gfeatures}.

5G offers high speed communication as an enhanced broadband service with speeds up to 10Gbps, extreme high capacity of 10Tbps per square KM, ultra high density with 1 million nodes per square KM, while 5G wide area networks support speeds up to a maximum of 40Gbps with supporting higher frequency ranges (24–86 GHz). Similarly, it supports ultra high user mobility of up to 500 KM per hour. The device to device communication under 5G IoT enables to connect billions of devices that revolutionize the modern industries including applications at agriculture, medical, manufacturing, business communication etc. Having latency less than a millisecond enables effective execution of mission critical applications operating in the real time. 5G is standardized to have extreme low energy consumption by IoT devices that supports battery life of more than 10 years. It has deep wireless coverage where the challenging locations can effectively be covered. SDN/NFV with its feature of network slicing is the backbone of 5G networks in the WAN for high speed programmable, flexible, scalable, and efficient network management. To identify massive IoT devices in the cyber physical systems (CPS) that are connected, controlled and communicated, IPv6 addressing is the current need after the depletion of IPv4 address space. Hence, IPv6 addressing and SDN are regarded as the backbone of 5G network to meet the requirements defined.

\subsection{IPv6, SDN, and the 5G}
\begin{figure}[ht!]
    \centering
    \includegraphics[width=0.8\textwidth]{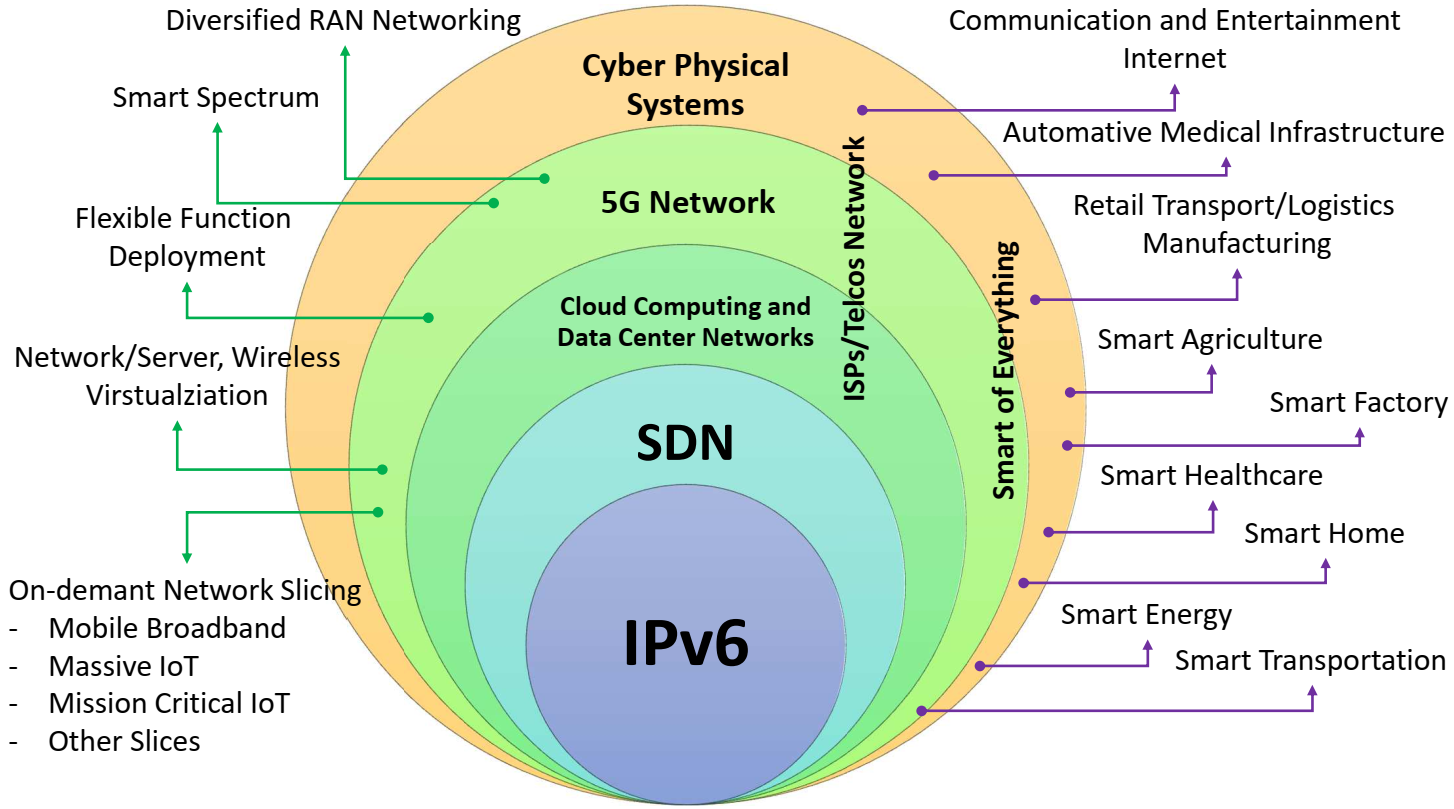}
    \caption{Layered view of IPv6, SDN, and the 5G network}
    \label{Fig:5GSDNIPv6}
\end{figure}

Figure \ref{fig:FutureNetworkParadigm} is revisited with Figure \ref{Fig:5GSDNIPv6}, where the latest smart network is considered with the incorporation of 5G. From core networking management and operations with SDN and IPv6 towards end access smart services to customers and enterprises in the CPS. The sufficiency of IP addresses provisioned by IPv6 allows for everything smart and communicable with the evolvement of IoT and WSN. Similarly, the programmability feature of SDN helps to introduce smartness on every devices. Figure \ref{Fig:5GSDNIPv6} shows the amalgamation of networking paradigms and their operations with services into layers. IPv6 and SDN are interrelated, because IPv6 deals with routing and addressing in the IP layer, while SDN deals with the controlling of networking operations as a networking management layer. Those technologies, which are recognized as network operation layer are operated by service providers. The customer services to be provided by ISPs and Telcos are the service layer activities. SDN and IPv6 have become increasingly important in the modern network environment as a result of the paradigm shifts in mobile communications brought about by the conceptualization and execution of 5G. 5G includes wired/wireless network and server virtualization, high-speed communications at densely populated smart devices with ultra-low latency for real-time mission critical applications, energy-efficient smart network deployment with efficient radio access network (RAN) design, and smart spectrum management and implementation. As a result, availability of 5G, SDN, and IPv6 is largely related to cloud computing, fog computing, and edge computing, from core network to end-access network service provisioning.

The next generation core transport network is conceptualized by transport-SDN (T-SDN), while the wireless network virtualization \cite{feng2015effective,kitindi2017wireless} is the part of 5G network and services. The high capacity, high density, massive IoT communications is possible only via the IPv6 addressing. To minimize the higher energy consumption by massively connected IoT devices, low power wireless personal area network (LoPAN) is designed for IPv6 as the standards of 6LoWPAN, which defines encapsulation and header compression for IPv6 packet communication in IP based sensor network environment \cite{mulligan20076lowpan}.  

The end-access network, which interacts with CPS, provides the fully 5G based services highlighted in Figure \ref{Fig:5GSDNIPv6} and fulfill the 5G network requirements considering the features listed in Figure \ref{Fig:5Gfeatures}.

\begin{figure}[ht!]
    \centering
    \includegraphics[width=0.7\textwidth]{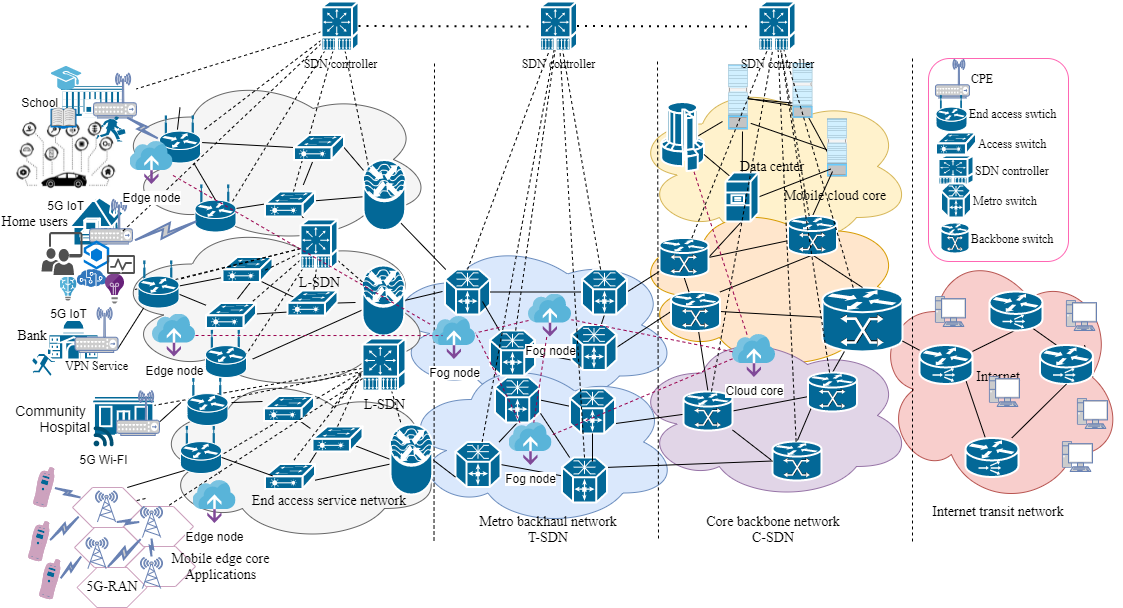}
    \caption{5G based SDN enabled IPv6 network aggregated from end-access to core backbone network}
    \label{Fig:5GAggregatedtedNetwork}
\end{figure}

Figure \ref{Fig:5GAggregatedtedNetwork} depicts the use case scenario of SDN enabled 5G aggregated network. It shows the hierarchy from core to end-access network, where computational intelligence in the end-access/IoT/WSN network can be applied by edge computing nodes then to FOG computing nodes and finally to cloud centers. Core network is controlled by Core SDN controller (C-SDN), while the transport network management form cloud to edge computing is controlled by Transport-SDN (T-SDN). For local network or end-end access network control and management, local-SDN (L-SDN) controllers are to be deployed. 

\section{Summary and Future Works}
\label{sec:Summary}
This paper presents the implementation of latest networking paradigms viz IPv6 and SDN in the service provider networks. Individual and joint migration methods and practices are discussed. Joint migration to SDN enabled IPv6 network is analyzed to avoid the issues and challenges of network migration for CapEX and OpEX optimization in the network migration. SDN and IPv6 networking standards and services are becoming the part of next generation wireless network called 5G networks. Migration from 4G to 5G network and services together with the integration of SDN and IPv6 networking paradigms are considered as future works.  

\section*{Acknowledgements}
This work is a part of PhD thesis carried out by Dr. Babu R. Dawadi \cite{dawadiThesis2021sodip6}. We acknowledge following organizations for this research support:
(1) UGC-Nepal (grant id: FRG/74\_75/Engg-1), (2) NTNU (Norwegian University of Science and Technology) under Sustainable Engineering Education Project (SEEP) financed by EnPe, (3) NAST-Nepal, and (4) UPV-ERASMUS+ KA107.

\bibliographystyle{unsrt}  
\bibliography{references}  





\end{document}